\documentclass[repreprint,onecolumn,nofootinbib,superscriptaddress,showkeys,citeautoscrip]{revtex4}
\pdfoutput=1
\usepackage[usenames,dvipsnames]{color}
\usepackage{ifpdf}
\ifpdf
  \usepackage{feynmf} 
\else
   \usepackage{feynmp} 
\fi
\usepackage{graphicx}
\usepackage{amsmath}
\usepackage{amsfonts}
\usepackage{amssymb}
\usepackage{dsfont}
\usepackage{dcolumn}
\usepackage{bm}
\usepackage{slashed}
\usepackage{pstricks}

\usepackage{slashed}
\usepackage{multirow}
\usepackage{array}
\usepackage{rotating}
\usepackage{xcolor}
\usepackage{epstopdf}
\usepackage{fancybox}

\newcolumntype{x}[1]{
>{\centering}p{#1}}%

\newcommand{\GeV}      {~\mathrm{GeV}}

\def \cha{\widetilde{\chi}^{\pm}_1}

\newcommand{\beqn}{\begin{eqnarray}}
\newcommand{\eeqn}{\end{eqnarray}}
\newcommand{\be}{\begin{equation}}
\newcommand{\ee}{\end{equation}}

\newcommand{\mathsym}[1]{{}}

\def \cha{\tilde{\chi}^{\pm}_1}

\def \n34{\tilde{\chi}^{0}_{3,4}}

\def \ta{\tilde{t}_1}

\def \sta{\tilde{\tau}_1}

\def\met100{\slashed{E}_T\geq 100 \GeV}

\newcommand{\gappeq}{\mathrel{\rlap {\raise.5ex\hbox{$>$}}
{\lower.5ex\hbox{$\sim$}}}}
\newcommand{\lappeq}{\mathrel{\rlap{\raise.5ex\hbox{$<$}}
{\lower.5ex\hbox{$\sim$}}}}

\usepackage{ulem,fancyvrb}
\def\MT{m_{\rm T}}
\def\met{\slashed{E}_{T}}
\def\meff{m_{\rm eff}}

\begin{document}

\title{Interpreting  the  First CMS and ATLAS SUSY Results}

\author{Sujeet~Akula}
\affiliation{Department of Physics, Northeastern University,
 Boston, MA 02115, USA}
 
\author{Ning~Chen}
\affiliation{C.N.\ Yang Institute for Theoretical Physics, 
Stony Brook University, Stony Brook, NY 11794, USA}

\author{Daniel~Feldman}
\affiliation{Michigan Center for Theoretical Physics,
University of Michigan, Ann Arbor, MI 48109, USA}

\author{Mengxi~Liu}
\affiliation{Department of Physics, Northeastern University,
 Boston, MA 02115, USA}

\author{Zuowei~Liu}
\affiliation{C.N.\ Yang Institute for Theoretical Physics, 
Stony Brook University, Stony Brook, NY 11794, USA}

\author{Pran~Nath}
\affiliation{Department of Physics, Northeastern University,
 Boston, MA 02115, USA}

\author{Gregory~Peim}
\affiliation{Department of Physics, Northeastern University,
 Boston, MA 02115, USA}


\begin{abstract}
The CMS and the ATLAS Collaborations have recently reported on the search for supersymmetry
with 35 pb$^{-1}$ of data and have put independent  limits on the parameter space of 
the supergravity unified model with universal boundary conditions at the GUT scale for soft 
breaking, i.e., the mSUGRA model. We extend this study by examining other regions of 
the mSUGRA  parameter  space in $A_0$ and $\tan\beta$. Further, we contrast the reach of 
CMS and ATLAS with 35~pb$^{-1}$ of data with the indirect constraints, i.e., the constraints 
from  the Higgs boson mass limits, from flavor physics and from the dark matter limits from 
WMAP. Specifically  it is found that a significant part of the parameter space excluded by 
CMS and ATLAS  is essentially already excluded by the indirect constraints 
and the fertile region of parameter space has yet  to be explored. We also emphasize 
that  gluino masses as low as 400 GeV but for squark masses much larger than the gluino mass
remain unconstrained and further that much of the hyperbolic branch of radiative electroweak 
symmetry breaking, with low values  of the Higgs mixing parameter $\mu$, 
is essentially  untouched by the recent LHC analysis.
\end{abstract}

\maketitle


\section{Introduction}
A candidate model for new physics is the $N=1$ supergravity grand unified model~\cite{sugra}
which with  universal boundary conditions  for soft breaking at the unification scale is the model 
mSUGRA~\cite{sugra,hlw,ArnowittNath} (for reviews see ~\cite{sugraR,KaneFeldman,Hunt}) 
defined by the parameter space $m_0,m_{1/2},A_0,\tan\beta$ and the sign of $\mu$, as well as 
$M_G$ and $\alpha_G$ where $M_G$ is the grand unification scale and $\alpha_G$ is the 
common value of $\alpha_1, \alpha_2, \alpha_3$  ($\alpha_i= g_i^2/(4\pi)$ and $g_i$ is 
gauge coupling) for the gauge groups $U(1)\times SU(2)_L\times SU(3)_C$ at the unification scale. 
This model has recently been investigated at the LHC with R parity conservation, and constraints 
on the model have  been set with 35 pb$^{-1}$  of data by the CMS and ATLAS collaborations 
\cite{cmsREACH,AtlasSUSY,atlas0lep}. These works, therefore,  produce the first direct constraints 
on supergravity unified models at the LHC. Indeed the recent results of CMS and ATLAS  
\cite{cmsREACH,AtlasSUSY, atlas0lep}  are encouraging as they report to
surpass the  parameter space probed 
in previous \cite{FengGrivazNachtman} direct searches by LEP and by the Tevatron.

In this work we explore and interpret  further the recent data from  the LHC in the framework of 
mSUGRA. In our analysis we follow the techniques developed in Ref. \cite{land1}, and we study 
the parameter space under full simulation of the 7 TeV standard model backgrounds 
\cite{Lessa,Peim,Peim2,KaneDF,Wacker}. The ATLAS analysis produces a reach which is more 
stringent that the one from CMS and thus in our analysis we utilize the cuts used by ATLAS. 
For the case $A_0=0$ and $\tan\beta=3$ our results are in conformity with the ATLAS  results 
but we explore other regions of the parameter space in $ A_0-\tan\beta$ plane and discuss the 
reach plots for these. We also carry out a detailed comparison of the constraints arising from the  
CMS  and ATLAS reach plots~\cite{cmsREACH,AtlasSUSY,atlas0lep} vs the constraints arising 
from the Higgs mass lower limits, from flavor physics and specifically from 
$\mathcal{B}r\left(b\to s\gamma\right)$  experimental bounds, and from the WMAP relic 
density constraints on  dark matter.


\section{Reach plots with {\boldmath $35~{\lowercase{\rm pb}}^{-1}$} of data}

The ATLAS collaboration has released two analyses, one with $1$~lepton~\cite{AtlasSUSY} and 
the other with $0$~leptons~\cite{atlas0lep}~both of which are considered in our analysis. 
For the $1$~lepton analysis we follow the selection requirements that ATLAS reports in 
\cite{AtlasSUSY}. The preselection requirements for events are that a jet must 
have $p_{T}>20\GeV$ and $\left|\eta\right|<2.5$, electrons must have $p_{T}>20\GeV$ 
and $\left|\eta\right|<2.47$ and muons must have $p_{T}>20\GeV$ and $\left|\eta\right|<2.4$. 
Further, we veto the ``medium" electrons\footnote{See \cite{atlasTDR} for a definition 
of ``loose", ``medium" and ``tight" electrons} in the electromagnetic calorimeter transition 
region, $1.37<\left|\eta\right|<1.52$. An event is considered if it has a single lepton 
with $p_{T}>20\GeV$ and its three hardest jets have $p_{T}>30\GeV$, with the leading 
jet having $p_{T}>60\GeV$. The distance, 
$\Delta R=\sqrt{\left(\Delta \eta\right)^2+\left(\Delta \phi\right)^2}$, between each 
jet with the lepton must satisfy $\Delta R\left(j_{i},\ell\right)>0.4$, and events are rejected if 
the reconstructed missing energy, $\met$, points in the direction of any of the three leading 
jets, $\Delta \phi \left(j_{i},\met\right)>0.2$. Events are then classified into $2$ channels, 
depending on whether the lepton is a muon or an electron. These are then further classified 
into four regions based on the missing energy and $\MT$ cuts, where 
we reconstruct the missing transverse momentum using the selected lepton plus 
jets with $p_{T}>20\GeV$ and $\left|\eta\right|<4.9$ following ATLAS analysis, and 
$\MT=\sqrt{2p_{T}\left(\ell\right)\met\left(1-\cos\left(\Delta\phi\left(\ell,\met\right)\right)\right)}$ 
is the transverse mass between the lepton and the missing transverse momentum vector. 
The four regions alluded to above are labeled the ``signal region", the ``top region'', 
the ``W region'' and the ``QCD region''. For the ``signal region" events were required to 
pass the additional cuts of 
$\MT>100\GeV$, $\met>125\GeV$, $\met>0.25\meff$ and $\meff>500\GeV$. 
Here the effective mass, $\meff$, is the scalar sum of the missing energy with the $p_{T}$'s 
of the selected visible objects (in this case the lepton and the $3$~jets).
The number of events were then compared to the $95\%~{\rm CL}$ upper bounds that 
ATLAS found ($N_{e}<2.2$~events and $N_{\mu}<2.5$~events)~\cite{AtlasSUSY}. 
The ``top region" and ``W region" are defined by events with $30\GeV<\met<80\GeV$ 
and $40\GeV< \MT< 80\GeV$, where the ``top region'' requires at least one of the three 
hardest jets to be $b$-tagged and the ``W region'' requires none of the three hardest jets to 
be $b$-tagged. The ``QCD region" was required to have $\MT,\met <40\GeV$ and was 
purely data driven. For our analysis events were rejected if they contaminated the three control regions.
Using the standard model background from \cite{Peim} we reproduced the ATLAS results.

\begin{figure*}[h!]
\includegraphics[scale=0.26]{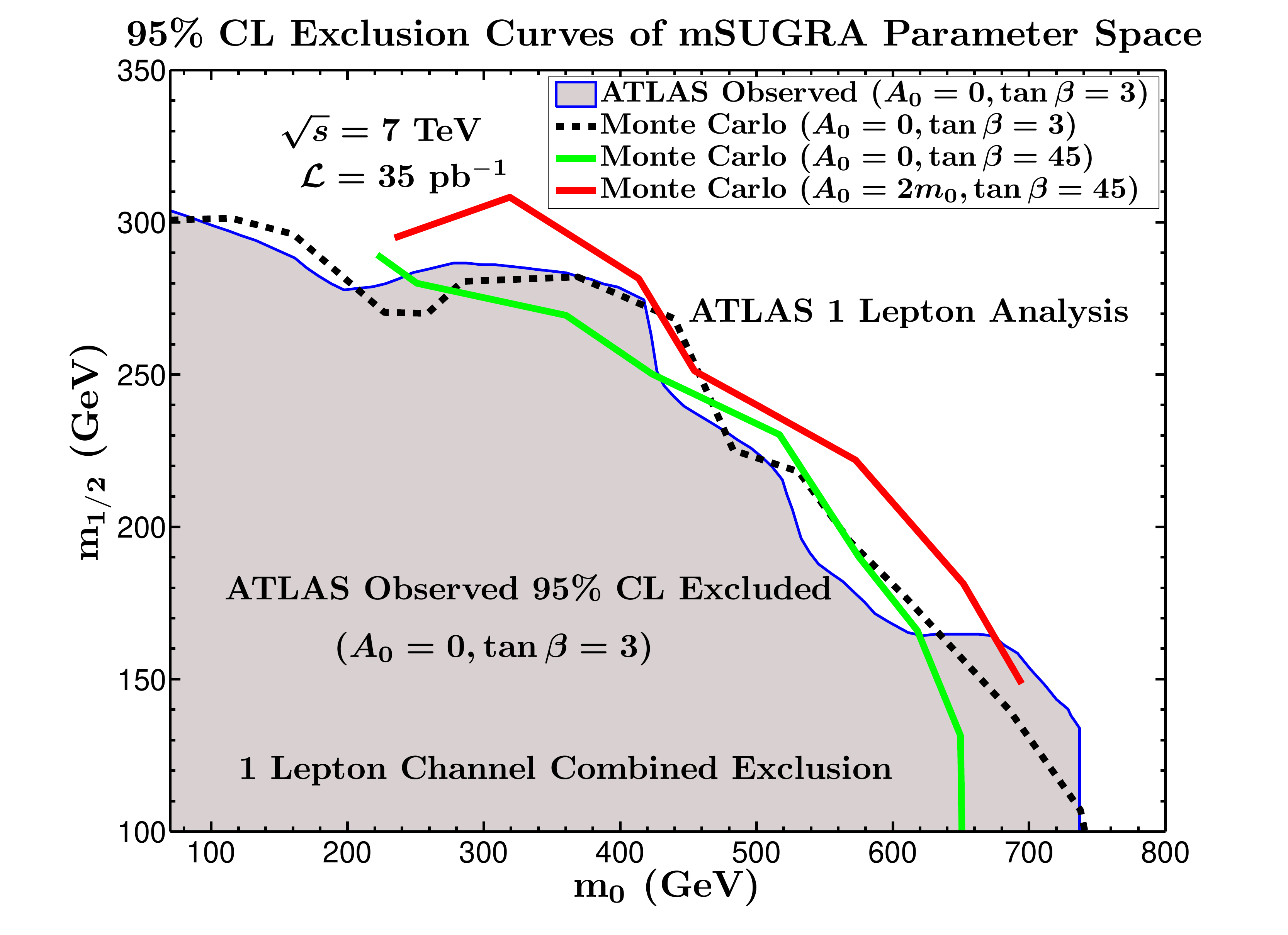} \hspace{-.1cm}
\includegraphics[scale=0.26]{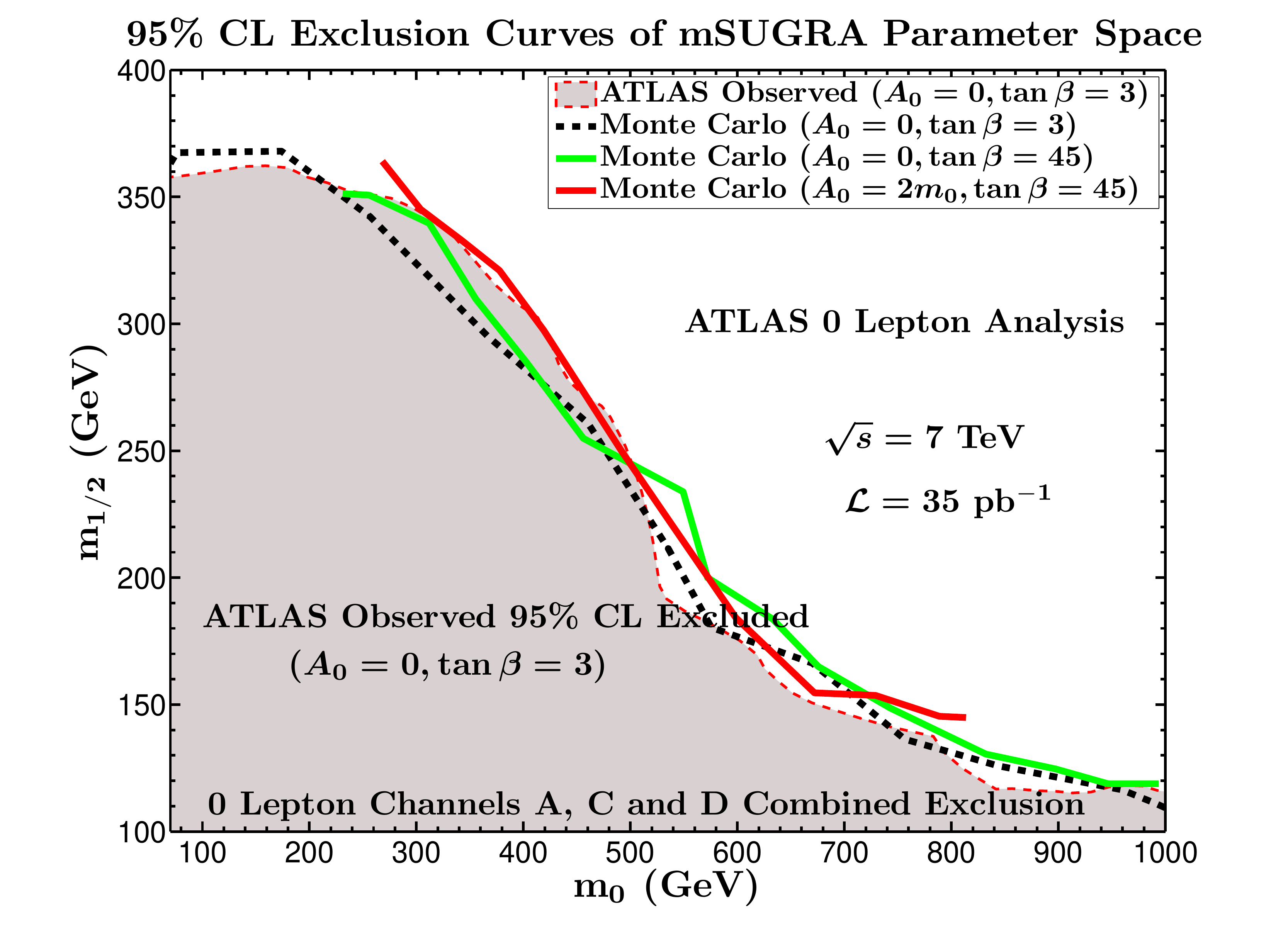}
\caption{\label{ReachCombine} (color online)
Left: Reach plot with $35$ pb$^{-1}$ of integrated luminosity using the ATLAS cuts 
\cite{AtlasSUSY} \cite{atlas0lep} with different $\tan\beta$ and $A_0$: $A_0=0$ 
and $\tan\beta=3$ (dashed line); $A_0=0$ and $\tan\beta=45$ (solid green line);
$A_0=2m_0$  and $\tan\beta=45$ (solid red line). For comparison we give the ATLAS 
observed limit ($A_0=0$ and $\tan\beta=3$) (solid blue line). Right:  Reach plot with 
35~pb$^{-1}$ of integrated luminosity of data using the ATLAS $0$ lepton cuts.
For comparison we give the ATLAS observed limit (red dashed line).}  
\label{atlas1}
\end{figure*}

For the $0$~lepton analysis we follow the selection requirements that ATLAS reports 
in~\cite{atlas0lep} where the pre-event selection is the same as for the $1$~lepton case 
except that leptons are identified to have $p_{T}>10\GeV$. Here the events are classified 
into 4 regions ``A", ``B", ``C" and ``D"; where regions A and B have at least $2$~jets and 
regions C and D have at least $3$~jets. When referring to different cuts in these regions we 
define cuts on the ``selected" jets to mean that the bare minimum number of jets in this region 
must satisfy the following requirement: For  regions A and B   ``selected" jets mean that they 
are the first two hardest jets and for regions C and D   ``selected"  jets mean that they are the 
first three hardest jets. Events are required to have $\met>100\GeV$ and the selected jets 
must each have $p_{T}>40\GeV$ with the leading jet $p_{T}>120\GeV$. As in the case 
with $1$~lepton, events are rejected if the missing energy points in the direction of any of the 
selected jets, $\Delta \phi \left(j_{i},\met\right)>0.4$, where $i$ is over the selected jets. 
Region A requires events to have $\met>0.3\meff$ and $\meff>500\GeV$ and  regions C 
and D require events to have $\met>0.25\meff$ with region C requiring $\meff>500\GeV$ 
and region D requiring $\meff>1~{\rm TeV}$. In this case $\meff$ is defined in terms of 
selected jets, i.e. for regions A and B it is the scalar sum of the first two hardest jets and for 
regions C and D it is the scalar sum of the first three hardest jets. For the analysis here we 
do not apply the cut for region B, i.e.  $m_{\rm T 2}>300\GeV$, since the models excluded 
in this region are already excluded in region D~\cite{atlasWeb}.

Following the framework of the ATLAS Collaboration \cite{AtlasSUSY} we have carried 
out a set of three parameter sweeps in the $m_0-m_{1/2}$ plane 
taking $m_{1/2}\leq 500~{\rm GeV}$ and $m_{0}\leq1~{\rm TeV}$.  Two of the 
parameter sweeps were a $10~{\rm GeV}\times10~{\rm GeV}$ grid scan in 
the $m_0-m_{1/2}$ plane having  a fixed universal trilinear parameter, $A_0=0$, 
and fixed $\tan\beta$; one set with $\tan\beta=3$ and the other with $\tan\beta=45$. 
A third parameter scan was done with $A_0=2 m_0$ and $\tan\beta =45$. Throughout 
the analysis  we take $\mu>0$ and $m_{\rm top}^{{\rm pole}} = 173.1\,{\rm GeV}$. 
For the simulation of the 
mSUGRA models, renormalization group evolution and computation of the physical masses 
of the sparticles was performed using {\tt SuSpect}~\cite{suspect} and we implement 
both {\tt MadGraph} and {\tt Pythia} for event generation \cite{mad,pyth}. A comparison 
of our reach to the reach done by the ATLAS Collaboration is shown in Fig.(\ref{atlas1}).


\begin{figure*}[t!]
   \begin{center}
   \includegraphics[scale=0.33]{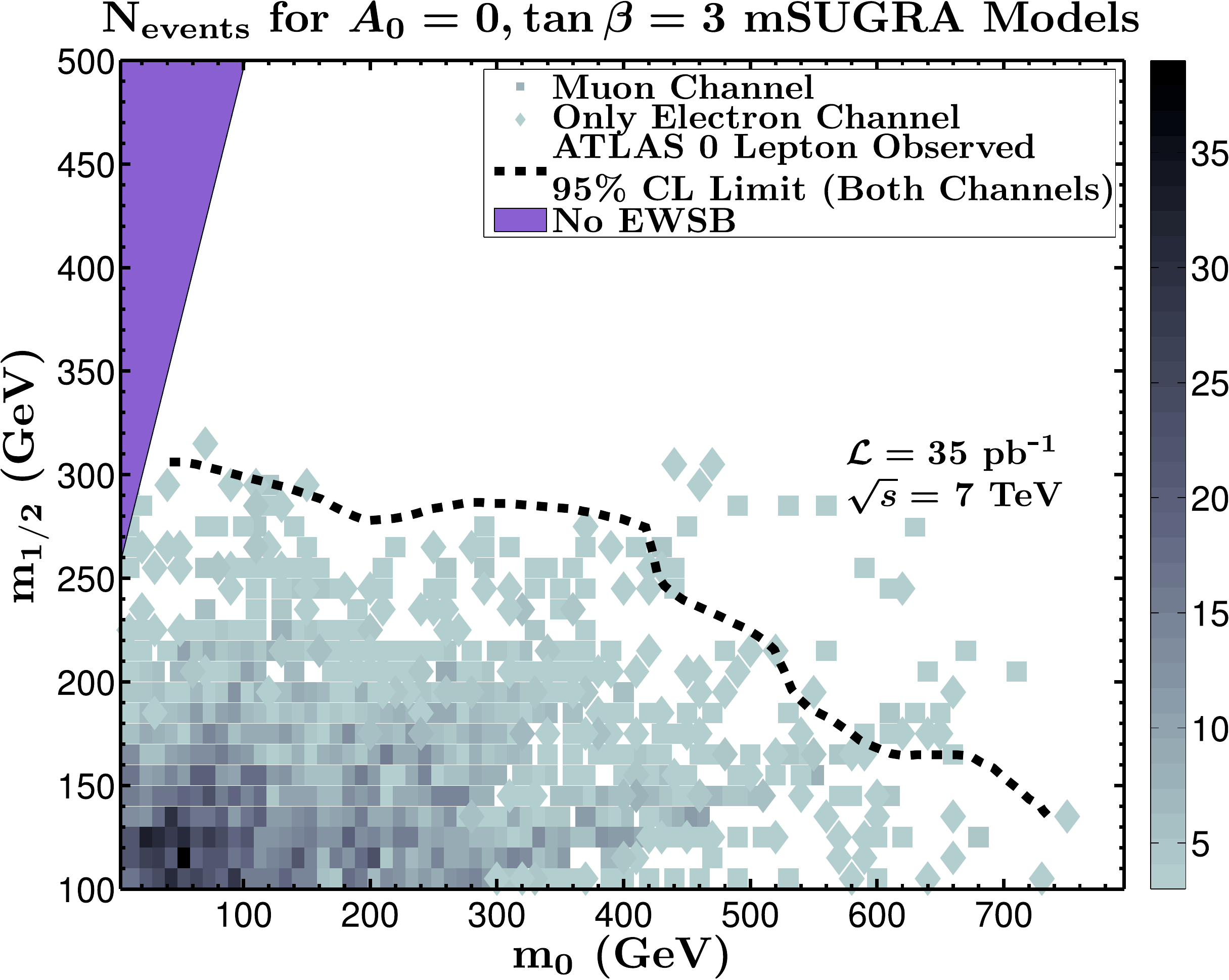}
   \includegraphics[scale=0.33]{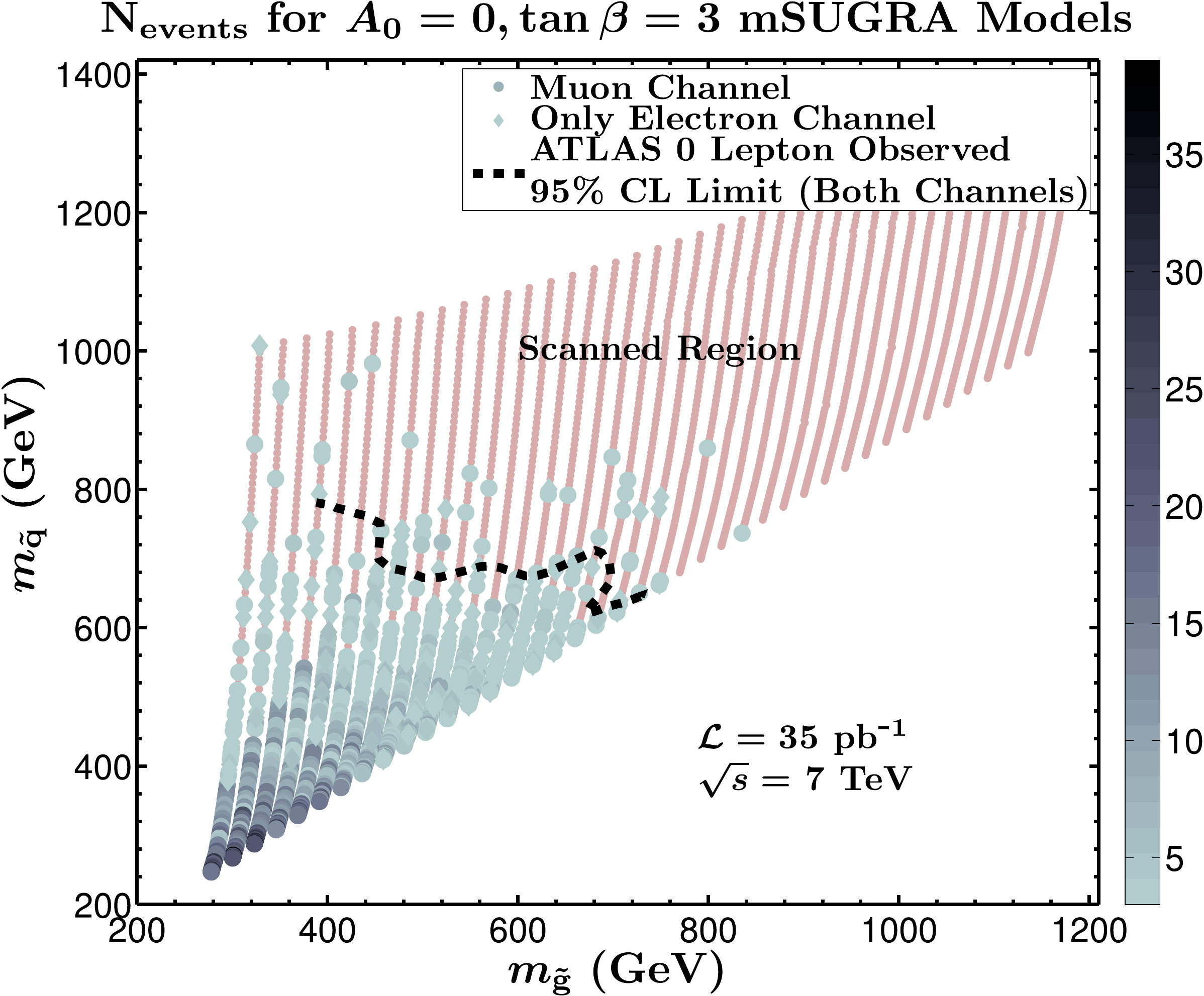}  \hspace{.25cm}    
   \includegraphics[scale=0.33]{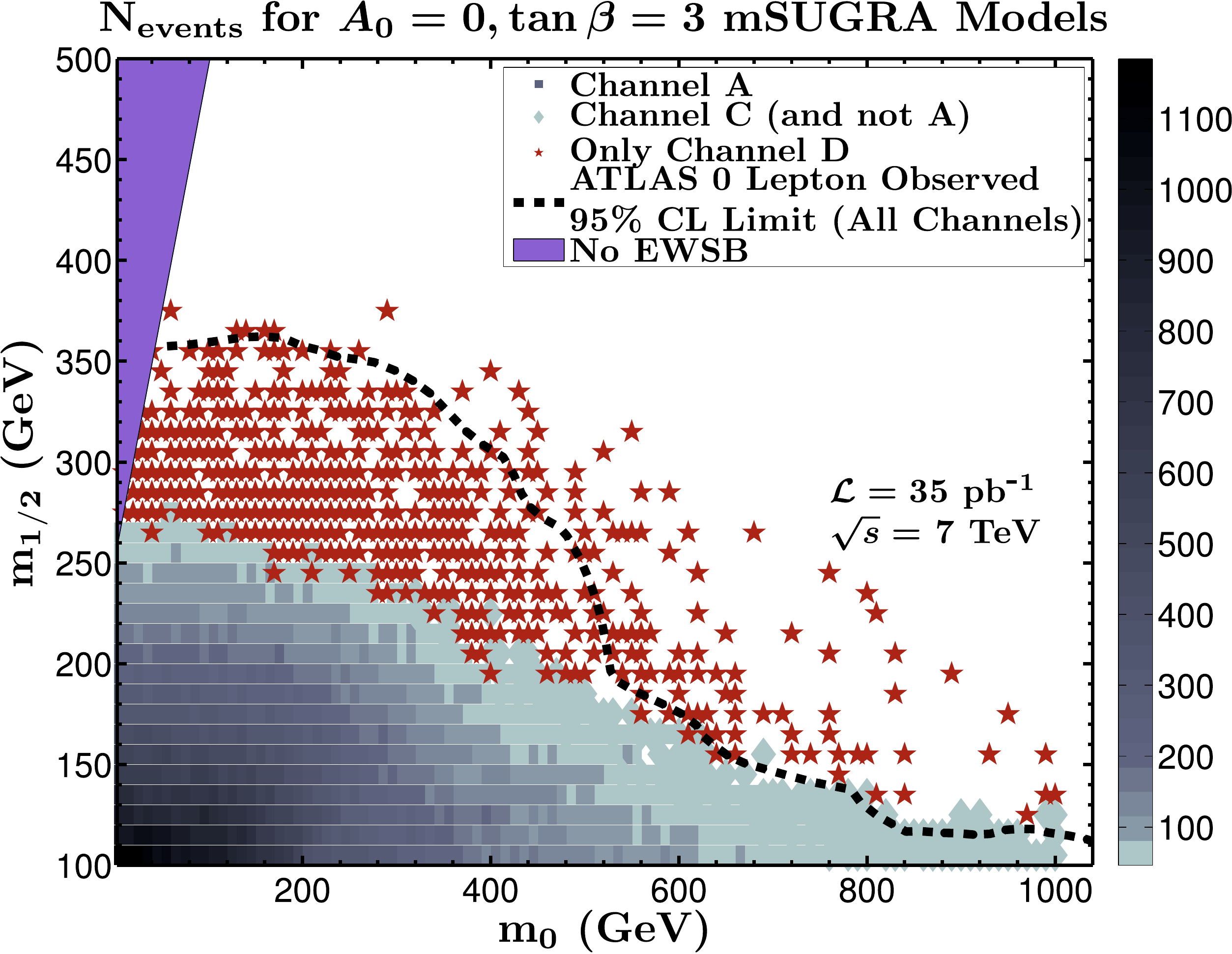}
   \includegraphics[scale=0.33]{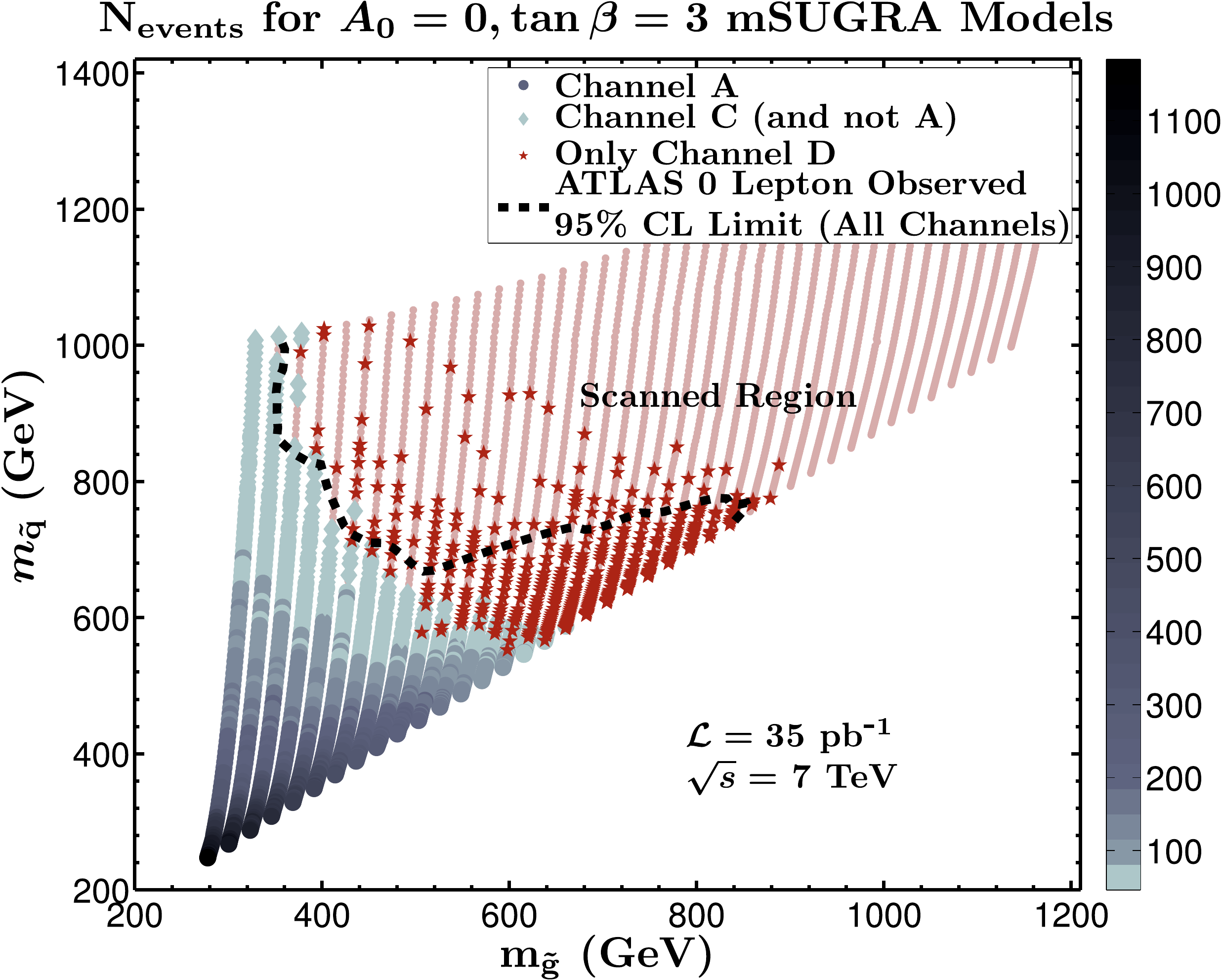}
\caption{\label{atlasEx} (color online) Top left panel:  Number of signal events in the 
$m_0-m_{1/2}$ plane for the case $A_0=0, \tan\beta =3$ using the $1$~lepton ATLAS 
cuts in the $m_0-m_{1/2}$ plane. 
The dark areas correspond to number of events  greater than $2$ with the actual 
numbers indicated along the vertical line to the right while the white areas are filled with models 
but have number of events less than 2. 
Top right panel: Same as the left panel except that the 
plot is $m_{\tilde g}(\rm gluino)-m_{\tilde q}(\rm squark)$ mass plane for the lightest squark 
of the first 2 generations. The square region in the left panel becomes squeezed into the polygon-like 
region in the physical mass plane in the right panel. One may note that the ATLAS constraints do 
not rule out a low mass gluino on the scale of order 400 GeV for heavy squarks. 
Bottom left panel: 
The same as the top left panel except that the analysis is done using $0$ lepton ATLAS cuts. 
Bottom right panel: Same as the top right panel except that the analysis is done using the $0$ lepton 
ATLAS cuts. The (red) stars correspond to channel  D.
In channel D we find maximally 51 events over the space scanned
after a requirement that the number of events be at least 15 before cuts.
However, when only considering models not already excluded by channels A and C,
the number of events in channel D is maximally 18.
}
\end{center}
\end{figure*}

In Figure~\ref{atlasEx} we plot the number of signal events  for electrons in the $m_0-m_{1/2}$ plane
where the reach plot from ATLAS is also exhibited and where the ATLAS reach plot corresponds to the
number of observed events and those that have a larger number predicted by the model. 
For the 1 lepton analysis, we first present the models excluded by the muon channel, 
colored by $N^\mu_{\rm events}$
(indicated by squares). Next, we overlay from the remaining models, those that have been excluded by the electron channel, and colored by $N^e_{\rm events}$ (indicated by diamonds). Similarly for the 0 lepton analysis, we begin with models excluded by channel A, colored by $N^A_{\rm events}$ (indicated by squares); overlay models excluded by C (but not A) and colored by $N^C_{\rm events}$ (indicated by diamonds). Next, we overlay models excluded by channel D alone in a single color (stars), as $N^D_{\rm events}$ are not comparable with $N^A_{\rm events}$ or $N^C_{\rm events}$.
We also show
the number of signal events for electrons in  the $m_{\tilde g}-m_{\tilde q}$ plane. An ATLAS reach 
curve is also exhibited. 

The upper left panel of Fig.(\ref{atlasEx}) gives us a more quantitative description of the electron and
muon channels in putting constraints on the $m_0-m_{1/2}$ parameter space with $35$ pb$^{-1}$ of data.
As expected the largest number of single $e$ and $\mu$ events arise at low mass scales, i.e., for 
low  values of $m_0$ and of $m_{1/2}$ and the number of signal events decrease and we approach the
boundary after which they fall below 2 for the 1 lepton ATLAS analysis.  It is also instructive to examine the signal events in the gluino-squark mass plane where the squark mass corresponds to the average first two generation 
squark mass.  This is done in the upper right panel of  Fig.(\ref{atlasEx}). Here the polygon shape of the region 
is a simple mapping of the allowed parameter in the $m_0-m_{1/2}$ plane of the upper left panel. 
The plot is useful as it directly correlates  squark and gluino  model points that are either excluded or allowed
by the 1 lepton ATLAS analysis. The 0 lepton analysis of the lower panels in Fig.(\ref{atlasEx})
is very similar to the analysis of the upper panels except for different array of cuts.  
There is a general consistency in the analysis of the 1 lepton and the 0 lepton analysis, although 
the 0 lepton cuts appear more constraining as they appear to exclude a somewhat  larger region
of the parameter space. Together 
 the analysis of the upper and lower panels of Fig.(\ref{atlasEx}) gives us a more analytical understanding 
 of the relative strengths of the 1 lepton and 0 lepton cuts.


\section{Implications of Constraints}

\begin{figure*}[h!]
   \begin{center}
   \includegraphics[scale=0.128]{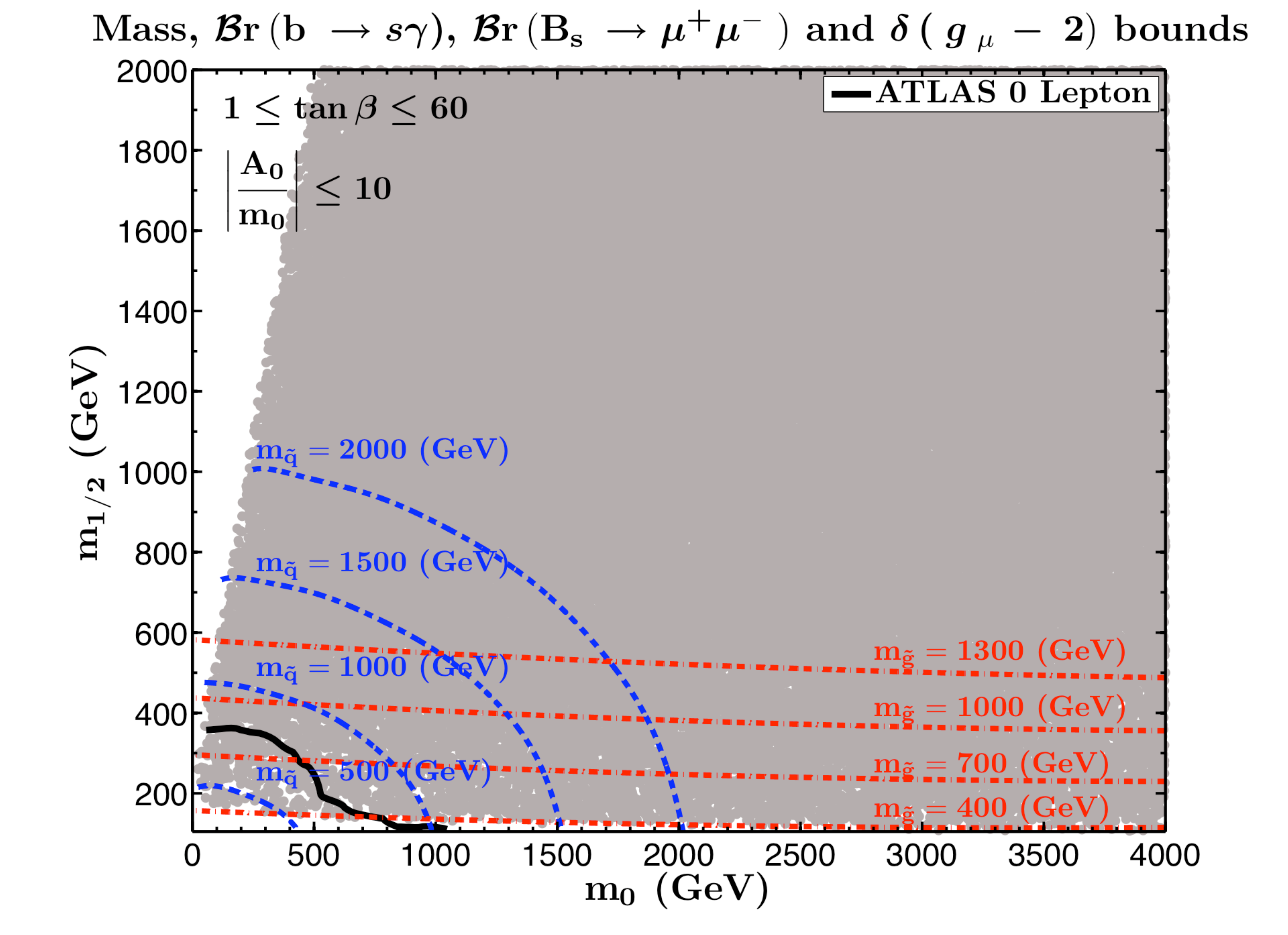}
   \includegraphics[scale=0.26]{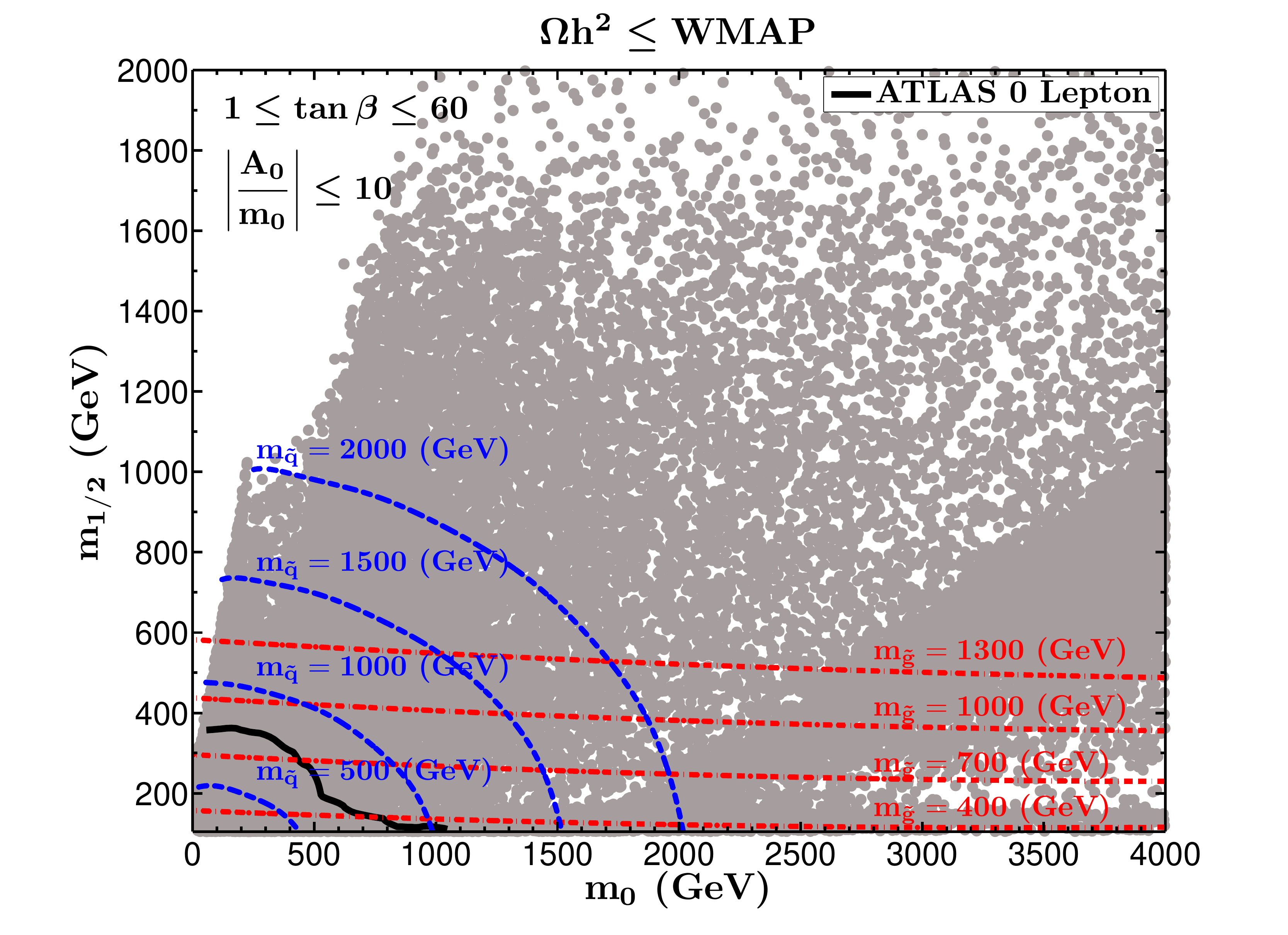}
     \includegraphics[scale=0.26]{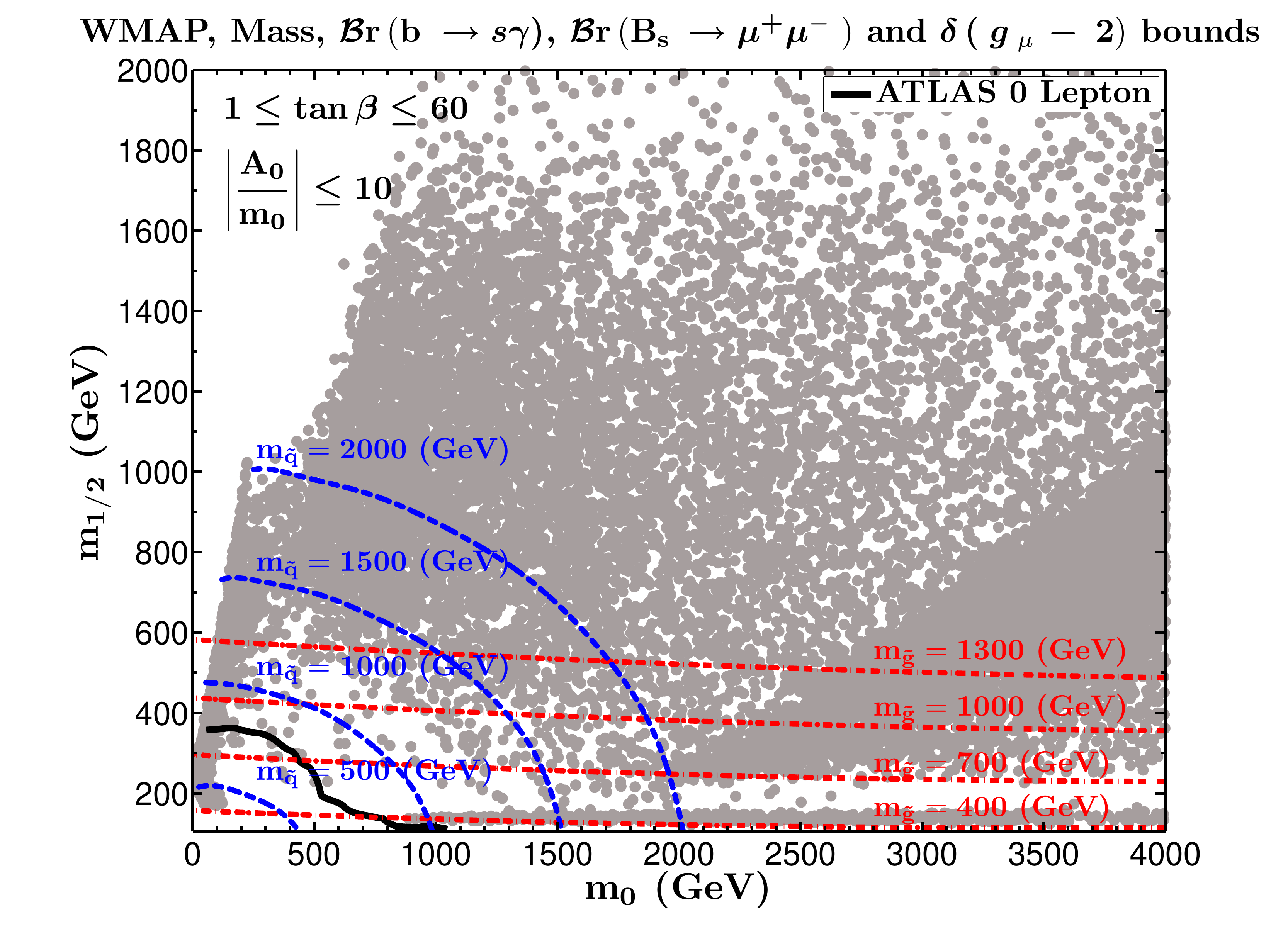}
   \includegraphics[scale=0.26]{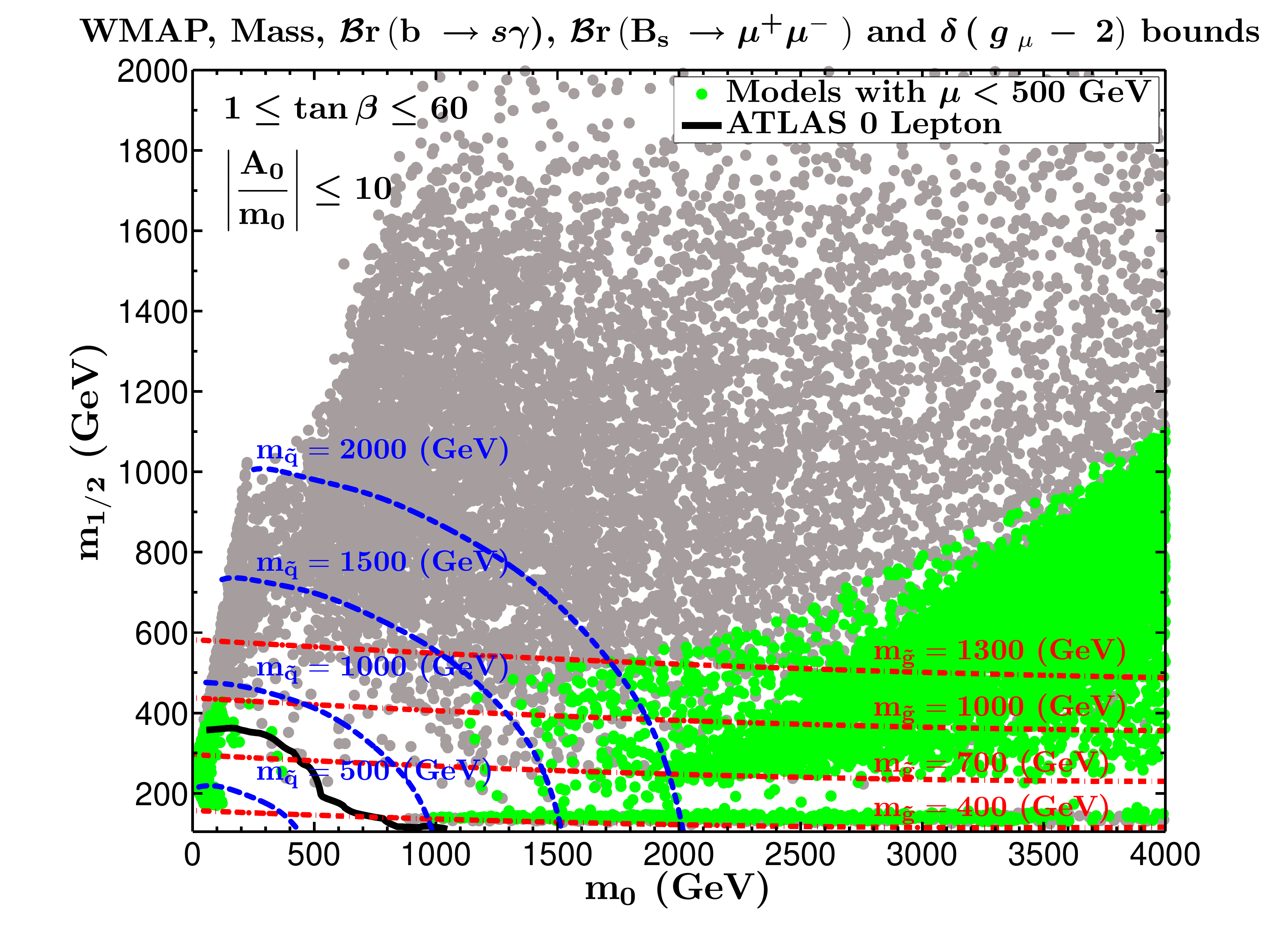}
\caption{  \label{total} (color online)  Upper left panel:
An exhibition of the allowed models indicated by grey (dark) dots in the $m_0-m_{1/2}$ plane 
when only flavor and collider constraints are imposed. The region excluded by ATLAS 
(as well as CMS)
lies below the thick black curve in the left hand corner. 
Upper right panel: same as the left upper panel except that only an upper bound on relic 
density of $\Omega h^2 \leq 0.14$ is imposed. 
Lower left panel: Same as the upper left panel except that  the relic 
density constraint as in the upper right panel is also applied. 
This panel exhibits that most of the parameter space excluded by ATLAS is already excluded
by the collider/flavor and relic density constraints. 
The dark region below the ATLAS
curve is the extra region excluded by ATLAS which was not previously excluded by the
indirect constraints.
Lower right panel: The analysis of this figure is similar to the lower left panel  
except that models with $|\mu|<500$ GeV are exhibited  in green.
   }
\end{center}
\end{figure*}

In the analysis of the reach plots experimental constraints were not imposed beyond those
that arise from the ATLAS analyses.  Next  we include these constraints 
and in our analysis we will consider 
 the larger parameter space when all four parameters
  $m_0, m_{1/2}, A_0, \tan\beta$ are varied.  
  In doing so, we  apply various constraints from searches
  on the sparticle mass limits, B-physics and from $g_{\mu} -2$.  Next we 
   explore the constraint from upper bound on the relic density from WMAP only, and then with combination 
   of all of the above.
   These indirect constraints were calculated using {\tt MicrOmegas}~\cite{micromegas}, with the Standard Model contribution in the ${\mathcal{B}r}\left(b\to s\gamma\right)$ corrected using the NNLO analysis of 
 Misiak~{\it et al}.~\cite{Misiak:2006zs,Chen:2009cw}.
 We now describe this more general analysis.
  In the upper left panel of Fig.(\ref{total})  we apply the following ``collider/flavor constraints" \cite{pdgrev} 
  $m_h > 93.5~\GeV$, $m_{\sta} > 81.9~\GeV$, $m_{\cha} > 103.5~\rm GeV$, and $m_{\ta} > 100~\GeV$, along with
   $\left(-11.4\times 10^{-10}\right)\leq
\delta \left(g_{\mu}-2\right) \leq \left(9.4\times10^{-9}\right)$, see~\cite{Djouadi:2006be},
${\mathcal{B}r}\left(B_{s}\to \mu^{+}\mu^-\right)\leq 4.7\times10^{-8}$ (90 \%  C.L.)~\cite{bsmumu}, and 
$\left(2.77\times 10^{-4} \right)\leq {\mathcal{B}r}\left(b\to s\gamma\right) \leq \left( 4.27\times 10^{-4}\right)$~\cite{bphys}. 
These collider/flavor constraints by themselves have an effect, but the effect is quite small in terms
of reducing the density of models that are already constrained by the ATLAS results.


We note that our scans of the parameter are very dense with  10$^6$ models
after EWSB alone.
In the $m_{0}-m_{1/2}$ plane the
collider/flavor cuts eliminate 12~\% of the models.  However
because $A_0$ and $\tan \beta$ are not fixed to specific values, but
are allowed to run
over their full natural ranges, a model point which is eliminated for say, large
$\tan \beta$ by   $b\to s \gamma$ or $B_{s} \to \mu^{+} \mu^{-}$ at a
specific point
in the $m_{0}-m_{1/2}$ plane can correspond to a model point with a smaller 
value of $\tan\beta$ for the same  $(m_{0},m_{1/2})$ which is not eliminated.Thus the $m_{0}-m_{1/2}$ plane
appears densely filled. This is contrary to what one would observe
for  fixed values of $(A_0,\tan \beta)$. For example, for
$(A_0,\tan \beta) = (0,45)$ the $b \to s \gamma$ constraint
would remove models at large $m_0$ up to close to 2 TeV and $m_{1/2}$
up to about 750 GeV. As another example,
for $(A_0,\tan \beta) = (0,3)$ (the space looked at by ATLAS, and
in the previous section) a strict limit of $m_h < 102 \GeV$ for light
CP even Higgs
removes all model points below the ATLAS limits. However because one is varying 
 $(A_0,\tan \beta)$  the area below the ATLAS limit is filled in this case.

  Continuing on  we  next consider the ``cosmological constraint" in the upper right panel of Fig.(\ref{total}) 
 where 
 we apply only an upper bound on the relic density of the thermally produced neutralino dark matter of $\Omega h^2 \leq 0.14$~\cite{WMAP}.
 The WMAP upper bound constraint removes
96.5~\% of the models alone, thus 
 this cosmological constraint is very severe eliminating
  a large fraction of models, but again the ATLAS constraints remain quite strong.

 Next we consider the ``combined  collider/flavor and cosmological constraints" and find that together these constraints are  generally much more severe than the ATLAS constraints. This is shown 
 in the lower left panel of Fig.(\ref{total}). Here models that were separately allowed by 
 previously known  collider/flavor constraints, and models that were separately allowed by just the upper bound
 from WMAP, are now eliminated under the imposition of the combined constraints.  
 There is, however, a new 
region that ATLAS appears to exclude above and beyond what the indirect
  constraints exclude and this region  is a  region for low $m_0$ and for $m_{1/2}$ around 350 GeV. 
  Thus it would require a larger integrated luminosity to move past the barren region,  which is above the ATLAS bound,
   to get into the fertile region of
 the parameter space,   where the fertile region is the area above the white patch in the lower panel of Fig.(\ref{total}).

Finally in  lower right panel of Fig.(\ref{total}) we show the value of $\mu$ 
(at the electroweak symmetry breaking scale) in the $m_{1/2}-m_0$ plane where $\mu$ is the Higgsino mass
 parameter that enters in the Higgs bilinear term  in the superpotential. The analysis is given under the ``combined constraints" discussed    in the lower left panel of Fig.(\ref{total}).
We  note that essentially all of the natural region of the parameter space corresponding to
small $\mu$, most of which lies close to the hyperbolic branch (Focus point) (HB/FP) \cite{HB} of radiative breaking of the 
electroweak symmetry or near the vicinity of the light CP even Higgs pole region \cite{Nath:1992ty}
 remains untouched by the CMS and LHC exclusion limits as illustrated in the lower right panel of  
  Fig.(\ref{total}) and
remains to be explored. 
Further, as pointed out in Ref. (1) of \cite{DFetal}, low mass gluinos as low as even 420 GeV in mSUGRA are allowed
for the region for  large $m_0$  where relic density can be satisfied on the  light CP even 
Higgs pole \cite{Nath:1992ty}. This  can be seen from  Fig.(\ref{total})  as the gluino and squark  
masses are exhibited in the plots. Along the 
Higgs pole region,
electroweak symmetry breaking can also be natural, i.e., one has a small $\mu$.  It is also seen that this  region  is 
not constrained by CMS and ATLAS since 
their limits taper off at large $m_0$ as $m_{\rm squark}$  gets heavy and the jets from squark production are depleted 
(see Ref. (1) of \cite{DFetal}).


\section{Conclusion}

The CMS  and ATLAS analyses on the search for supersymmetry are impressive in that 
with only 35~pb$^{-1}$  of data their reach plots already exceed those from CDF and 
D\O\ experiments at the Tevatron. Both CMS and ATLAS have given reach plots  in 
the $m_0-m_{1/2}$ plane for the case $A_0=0, \tan\beta =3$ with the ATLAS analysis 
presenting more stringent limits compared to CMS. Because of the more stringent 
 limits  from ATLAS we adopted the ATLAS cuts in our analysis presented in this work. 
In our analysis we find consistency with the 1 lepton and  0 lepton results of ATLAS for 
the case analyzed by ATLAS, i.e., $A_0=0, \tan\beta=3$. We have also investigated 
reach plots for other values of $A_0, \tan\beta$, i.e., $A_0=0, \tan\beta=45$ 
and $A_0=2, \tan\beta=45$.
Another interesting question explored in this work 
is a relative study of the constraints on the $m_0-m_{1/2}$ parameter space by the CMS 
and ATLAS experiments vs the constraints that arise from Higgs mass limits, flavor physics, 
and from the dark matter constraints from WMAP.  One finds that the current CMS and 
ATLAS limits are consistent with such constraints. Specifically a significant part of the 
parameter space excluded by the CMS and ATLAS 35 pb$^{-1}$ data is already excluded 
by the indirect constraints. We  emphasize  that  low gluino masses (even as low as 400 GeV)  
remain unconstrained in mSUGRA, and and this conclusion holds generically
for  other  high scale models of soft breaking,
for the case when the squark masses are significantly larger than the gluino mass. Of interest
to the model at hand, is that such
situation arises on the hyperbolic branch of radiative breaking of the electroweak symmetry 
where typically $\mu$ is relatively small, and the region is very dense in the allowed set of 
parameter points. Finally, we note that some  recent papers related to various  topics discussed 
in this work can be found in Ref.\cite{DFetal}.\\


\noindent {\it Acknowledgements:} 
PN and GP would like to thank Darien Wood for extended discussions.  DF would 
like to thank Gordy Kane, Katie Freese,  Aaron Pierce, and Ran Lu for discussions. 
NC and  ZL  thank  Jie Chen and Robert Shrock for discussions.  This research is  
supported in part by the Department of Energy (DOE) grant  DE-FG02-95ER40899,  
and the U.S. National Science Foundation (NSF) grants PHY-0757959, and PHY-0969739,  
and in addition by the NSF TeraGrid resources provided by National Center for 
Supercomputing Applications (NCSA) under grant number TG-PHY100036.\\



\begin{thebibliography}{999}


\bibitem{sugra}
  A.~H.~Chamseddine, R.~L.~Arnowitt and P.~Nath,
  Phys.\ Rev.\ Lett.\  {\bf 49}, 970 (1982);
  P.~Nath, R.~L.~Arnowitt and A.~H.~Chamseddine,
  Nucl.\ Phys.\  B {\bf 227}, 121 (1983).

  \bibitem{hlw}
  L.~J.~Hall, J.~D.~Lykken and S.~Weinberg,
  Phys.\ Rev.\  D {\bf 27}, 2359 (1983).

\bibitem{ArnowittNath}
  R.~Arnowitt and P.~Nath,
  Phys.\ Rev.\ Lett.\  {\bf 69}, 725 (1992).

  \bibitem{sugraR}
  P.~Nath,
  arXiv:hep-ph/0307123;
  P.~Nath, R.~L.~Arnowitt, A.~H.~Chamseddine,
  ``Applied N=1 Supergravity,'' World Scientific Singapore, 1984;
   H.~P.~Nilles,
  Phys.\ Rept.\  {\bf 110}, 1-162 (1984);
  L.~E.~Ibanez and G.~G.~Ross,
  Comptes Rendus Physique {\bf 8}, 1013 (2007).

  \bibitem{KaneFeldman}
G.~Kane et. al  ``Perspectives on supersymmetry. Vol.2,''
{\it  World Scientific (2010) 583 p}.

  \bibitem{Hunt}
  P.~Nath, B.~D. ~Nelson, D.~Feldman, Z.~Liu {\it et al.},
  Nucl.\ Phys.\ Proc.\ Suppl.\  {\bf 200-202}, 185 (2010).


\bibitem{cmsREACH}
  V.~Khachatryan {\it et al.}  [CMS Collaboration],
  arXiv:1101.1628 [hep-ex].

\bibitem{AtlasSUSY}
  T.~A.~Collaboration,
  arXiv:1102.2357 [hep-ex].

\bibitem{atlas0lep}
  T.~A.~Collaboration,
  arXiv:1102.5290 [hep-ex].



\bibitem{FengGrivazNachtman}
 For a review of LEP/Tevatron constraints see:  J.~L.~Feng, J.~F.~Grivaz and J.~Nachtman in Ref.~\cite{KaneFeldman}~and~
  Rev.\ Mod.\ Phys.\  {\bf 82}, 699 (2010).


  \bibitem{land1}
  D.~Feldman, Z.~Liu and P.~Nath,
  Phys.\ Rev.\ Lett.\  {\bf 99}, 251802 (2007);
  Phys.\ Lett.\  B {\bf 662}, 190 (2008);
  JHEP {\bf 0804}, 054 (2008);
  Phys.\ Rev.\  D {\bf 80}, 015007 (2009);
  Phys.\ Rev.\  D {\bf 81}, 095009 (2010).

    \bibitem{Lessa}
  H.~Baer, V.~Barger, A.~Lessa and X.~Tata,
  JHEP {\bf 1006}, 102 (2010).

\bibitem{Peim}
  B.~Altunkaynak, M.~Holmes, P.~Nath, B.~D.~Nelson, G.~Peim,
Phys.\ Rev.\  D {\bf 82}, 115001 (2010).

\bibitem{Peim2}
  N.~Chen, D.~Feldman, Z.~Liu, P.~Nath, G.~Peim,
  arXiv:1011.1246 [hep-ph],  Phys.\ Rev.\  D {\bf 83}, 035005 (2011);
 N. Chen et al., Phys.\ Rev.\ D {\bf 83}, 023506 (2011).


\bibitem{KaneDF}
 D.~Feldman, G.~Kane, R.~Lu and B.~D.~Nelson,
  Phys.\ Lett.\  B {\bf 687}, 363 (2010);
  G.~L.~Kane, E.~Kuflik, R.~Lu and L.~T.~Wang,
  arXiv:1101.1963 [hep-ph].



\bibitem{Wacker}
  E.~Izaguirre, M.~Manhart and J.~G.~Wacker,
  JHEP {\bf 1012}, 030 (2010);
    D.~S.~M.~Alves, E.~Izaguirre and J.~G.~Wacker,
  arXiv:1008.0407 [hep-ph].

\bibitem{mad}
  J.~Alwall {\it et al.},
  JHEP {\bf 0709}, 028 (2007);
  

\bibitem{pyth}
  T.~Sjostrand, S.~Mrenna and P.~Z.~Skands,
  JHEP {\bf 0605}, 026 (2006).


\bibitem{Barate:2003sz}
  R.~Barate {\it et al.}  [
  Phys.\ Lett.\  B {\bf 565}, 61 (2003).
\bibitem{tev}
    [CDF and D0 Collaboration],
  arXiv:1007.4587 [hep-ex].
  
   
  



\bibitem{micromegas}
  G.~Belanger, F.~Boudjema, P.~Brun, A.~Pukhov, S.~Rosier-Lees, P.~Salati and A.~Semenov,
  Comput.\ Phys.\ Commun.\  {\bf 182}, 842 (2011).

\bibitem{suspect}
  A.~Djouadi, J.~L.~Kneur and G.~Moultaka,
  Comput.\ Phys.\ Commun.\  {\bf 176}, 426 (2007).

\bibitem{susyhit}
 M.~Muhlleitner, A.~Djouadi and Y.~Mambrini,
  Comput.\ Phys.\ Commun.\  {\bf 168}, 46 (2005);
  A.~Djouadi, M.~M.~Muhlleitner and M.~Spira,
  Acta Phys.\ Polon.\  B {\bf 38}, 635 (2007).

\bibitem{pdgrev}
  K.~Nakamura {\it et al.} [ Particle Data Group Collaboration ],
  J.\ Phys.\ G {\bf G37}, 075021 (2010).

\bibitem{atlasWeb}
\url{http://atlas.web.cern.ch/Atlas/GROUPS/PHYSICS/PAPERS/susy-0lepton_01/}


\bibitem{Djouadi:2006be}
  A.~Djouadi, M.~Drees and J.~L.~Kneur,
  JHEP {\bf 0603}, 033 (2006).
  
\bibitem{bphys}
  E.~Barberio {\it et al.}  [Heavy Flavor Averaging Group],
  arXiv:0808.1297 [hep-ex].

\bibitem{Chen:2009cw}
  N.~Chen, D.~Feldman, Z.~Liu and P.~Nath,
  Phys.\ Lett.\  B {\bf 685}, 174 (2010).

  
  
  \bibitem{bsmumu}
  T.~Aaltonen {\it et al.}  [CDF Collaboration],
  Phys.\ Rev.\ Lett.\  {\bf 100}, 101802 (2008).

\bibitem{Misiak:2006zs}
  M.~Misiak {\it et al.},
  Phys.\ Rev.\ Lett.\  {\bf 98}, 022002 (2007).

\bibitem{WMAP} E.~Komatsu {\it et al.}  [WMAP Collaboration],
  Astrophys.\ J.\ Suppl.\  {\bf 192}, 18 (2011);
N.~Jarosik {\it et al.},
  [arXiv: 1001.4744 [astro-ph.CO]];
   D.~N.~Spergel {\it et al.} ,
   Astrophys.\ J.\ Suppl.\  {\bf 170}, 377 (2007);
    Astrophys.\ J.\ Suppl.\  {\bf 148}, 175 (2003).



\bibitem{atlasTDR}
  G.~Aad {\it et al.} [ The ATLAS Collaboration ],
%



\bibitem{HB}
  K.~L.~Chan, U.~Chattopadhyay and P.~Nath,
  Phys.\ Rev.\  D {\bf 58} (1998) 096004;
   R.~L.~Arnowitt and P.~Nath,
  Phys.\ Rev.\  D {\bf 46}, 3981 (1992);
 J.~L.~Feng,   K.~T.~Matchev and T.~Moroi,
  Phys.\ Rev.\ Lett.\  {\bf 84}, 2322 (2000);
    U.~Chattopadhyay, A.~Corsetti and P.~Nath,
  Phys.\ Rev.\  D {\bf 68}, 035005 (2003);
  H.~Baer,  C.~Balazs, A.~Belyaev, T.~Krupovnickas and X.~Tata,
  JHEP {\bf 0306}, 054 (2003);
    D.~Feldman, Z.~Liu and P.~Nath,
  Phys.\ Rev.\  D {\bf 78}, 083523 (2008);
S.~Cassel, D.~M.~Ghilencea, S.~Kraml, A.~Lessa, G.~G.~Ross,
    [arXiv:1101.4664 [hep-ph]].



\bibitem{DFetal}
  D.~Feldman, K.~Freese, P.~Nath, B.~D.~Nelson, G.~Peim,
    [arXiv:1102.2548 [hep-ph]];
       I.~Gogoladze, R.~Khalid, S.~Raza, Q.~Shafi,
    [arXiv:1102.0013 [hep-ph]];
B.~C.~Allanach,
    [arXiv:1102.3149 [hep-ph]];
  H.~K.~Dreiner, S.~Grab, T.~Stefaniak,
    [arXiv:1102.3189 [hep-ph]];
  S.~Scopel, S.~Choi, N.~Fornengo, A.~Bottino,
   [arXiv:1102.4033 [hep-ph]];
  O.~Buchmueller et.al., 
    [arXiv:1102.4585 [hep-ph]];
     M.~Guchait and D.~Sengupta,
  arXiv:1102.4785 [hep-ph];
P.~Bechtle et.al.,  
    [arXiv:1102.4693 [hep-ph]];
  D.~S.~M.~Alves, E.~Izaguirre, J.~G.~Wacker,
  [arXiv:1102.5338 [hep-ph]]; B.~C.~Allanach, T.~J.~Khoo, C.~G.~Lester and S.~L.~Williams,
  arXiv:1103.0969 [hep-ph].

\bibitem{Nath:1992ty}
  P.~Nath, R.~L.~Arnowitt,
  Phys.\ Rev.\ Lett.\  {\bf 70}, 3696-3699 (1993).
  [hep-ph/9302318].

\end{thebibliography}
\end{document}